\documentclass[pre,twocolumn,showpacs,epsfig]{revtex4}
\usepackage{graphicx}
\usepackage{graphics}
\usepackage{wrapfig}
\usepackage{dcolumn}
\usepackage{amssymb}
\usepackage{bm}
\usepackage{epsfig}
\usepackage{psfrag}
\usepackage{color}
\usepackage[utf8x]{inputenc}
\usepackage{ucs}
\usepackage{amsmath}
\usepackage{amsfonts}
\usepackage{amssymb}
\usepackage{graphicx}
\usepackage{dcolumn}
\usepackage{bm}
\usepackage{latexsym}
\usepackage{hyperref}

\usepackage{xcolor}
\hypersetup{
    colorlinks,
    linkcolor={blue!50!black},
    citecolor={blue!50!black},
    urlcolor={blue!80!black}
}

\def\deg{^{\circ}}

\def\3dots{\:\raisebox{-0.5ex}{$\stackrel{\textstyle.}{:}$}\:}
\def\beq{\begin{equation}}
\def\eeq{\end{equation}}
\def\bea{\begin{eqnarray}}
\def\eea{\end{eqnarray}}
\def\deg{^{\circ}}

\begin{document}

\title{{Trapping and sorting active particles: motility-induced condensation \& smectic defects}}
\author{Nitin Kumar$^{1,}$\footnote{These authors contributed equally to this work}$^{,}$\footnote{{Present address}: James Franck Institute, University of Chicago, Chicago, Illinois 60637, USA;\href{mailto:nitink@uchicago.edu}{nitink@uchicago.edu}}}
\author{Rahul Kumar Gupta$^{2,*,}$\footnote{\href{mailto:rahulkg@tifrh.res.in}{rahulkg@tifrh.res.in}}}
\author{Harsh Soni$^{1,3,*,}$\footnote{\href{mailto:harshiisc@gmail.com}{harshiisc@gmail.com}}}
\author{Sriram Ramaswamy$^{1,2,}$\footnote{\href{mailto:sriram@iisc.ac.in}{sriram@iisc.ac.in}}}
\author{A K Sood$^{1,}$\footnote{\href{mailto:asood@iisc.ac.in}{asood@iisc.ac.in}}}
 \affiliation{$^1$Department of Physics, Indian Institute of Science, Bangalore 560 012, India\\
 $^2$Tata Institute of Fundamental Research, Gopanpally, Hyderabad 500 107,  India\\$^3$School of Engineering, Brown University, Providence 02912, USA}

\begin{abstract}

We present an experimental realization of the collective trapping phase transition [Kaiser et al., \href{https://doi.org/10.1103/PhysRevLett.108.268307}{Phys. Rev. Lett. \textbf{108}, 268307 (2012)}], using motile polar granular rods in the presence of a V-shaped obstacle. We offer a theory of this transition based on the interplay of motility-induced condensation and liquid-crystalline ordering and show that trapping occurs when persistent influx overcomes the collective expulsion of smectic defect structures. In agreement with the theory, our experiments find that a trap fills to the brim when the trap angle $\theta$ is below a threshold $\theta_c$, while all particles escape for $\theta > \theta_c$. Our simulations support a further prediction, that $\theta_c$ goes down with increasing rotational noise. We exploit the sensitivity of trapping to the persistence of directed motion to sort particles based on the statistical properties of their activity.	

\end{abstract}
\maketitle

\section{Introduction} \label{sec:intro}
The interplay of directed energy transduction and interparticle interaction 
gives rise to a host of dramatic self-organizing effects in collections of 
active particles \cite{SriramAnnRev,SriramRMP}. Although the active matter 
paradigm was formulated to describe the living state, it is frequently more 
practical to study this class of materials using reconstituted systems  \cite{BauschNature,DogicNature} or by creating faithful imitations 
\cite{Sano,VJScience,VJthesis,dauchotPRL,DJDurian,NatCom}. Here we work with 
fore-aft asymmetric metal rods, millimeters in length, confined in a quasi-two-dimensional
geometry and rendered motile in the horizontal plane by vertical vibration. 
Such objects \cite{Sano,VJthesis,VJScience,NatCom}, which we shall refer to as active polar rods, are now a standard test bed for 
probing the collective \cite{NatCom} and single-particle \cite{NKPRL,NKPRE} statistical 
physics of self-driven matter. The dynamics of self-propelled particles is 
influenced by the shapes of boundaries and obstacles, often leading to clustering, trapping and rectification \cite{austin,kudrolli,capture,bristlebot}, raising questions of fundamental interest and opening up new possibilities for the processing of granular and colloidal material.  

Even in the absence of obstacles, active matter is prone to clustering, through motility-induced phase separation (MIPS) \cite{MIPS}, a consequence of persistent motion at high enough concentration. We show here that the presence of traps leads to an effect related to MIPS, at far lower area fraction.  

\begin{figure}[t]
		\includegraphics[width=0.5\textwidth]{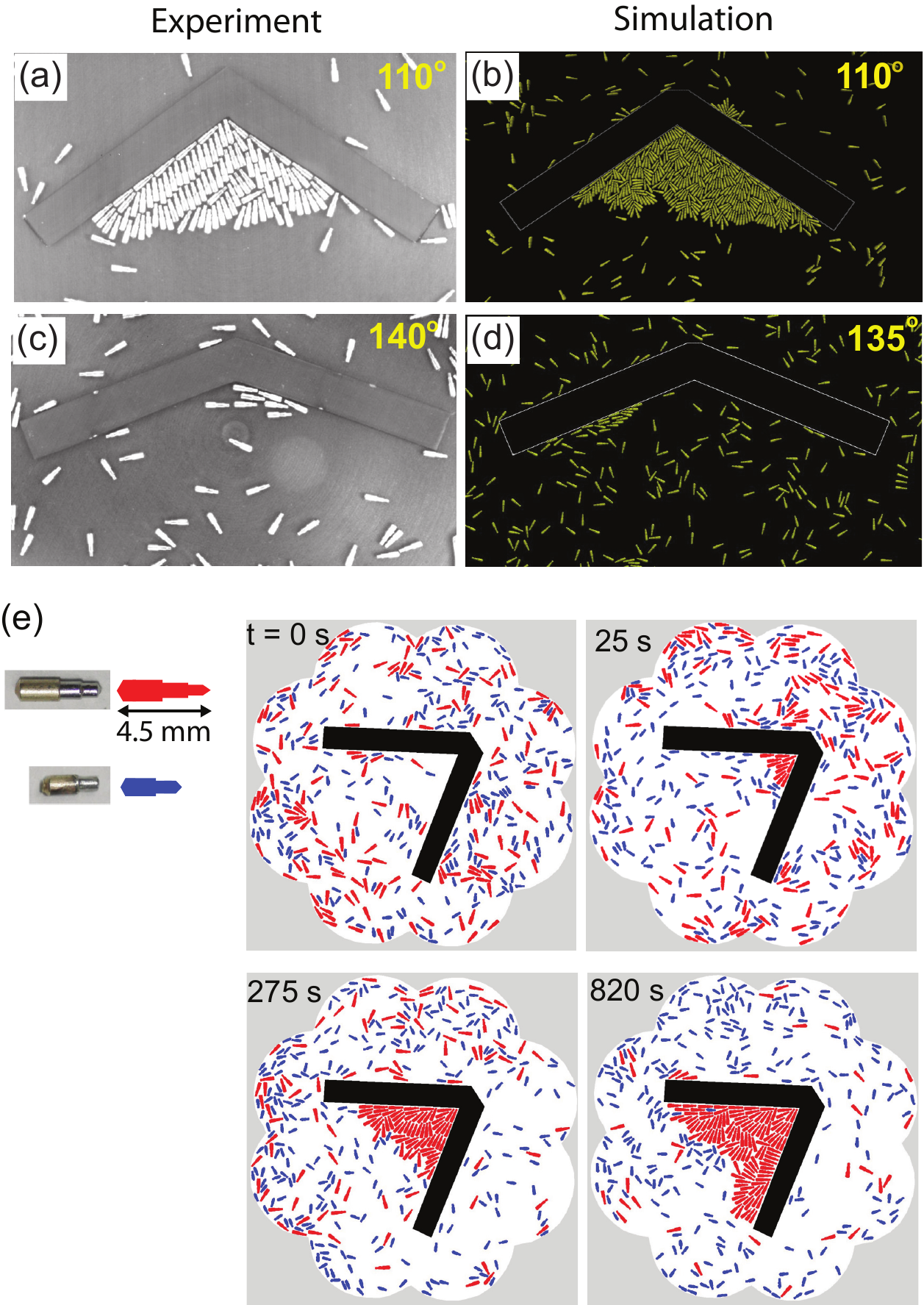}\\
		\caption{(a)-(d) Typical trapped and untrapped states in experiment and simulation. The angles are mentioned in yellow. In terms of rod length, the diameter of the outer circle of the flower-shaped geometry in experiments is 20, and the dimension of the periodic simulation box is 39. (e) A sequence of images showing separation of polar particles based on their activity with only the red particle getting trapped.}
		\label{results}
\end{figure}
In this paper we study an experimental realization of the collective trapping of active particle by Kaiser et al.\cite{capture}, in a system of motile granular rods.  
Through experiments, granular dynamic simulations, and analytical theory, we investigate the collective dynamics of mono- and bidisperse collections of such rods in the presence of a V-shaped trap. Our control parameters are the angle $\theta$ of the V and the nature of fluctuations in the motion of a single rod. In experiments the 
latter is governed by particle shape, while in simulations it can be varied continuously. 

Here is a summary of our results. (i) The polar rods undergo a phase transition to a collectively trapped state for $\theta < \theta_c$, a critical angle whose value is around $120^{\circ}$ for particles with strongly directed motion, as in \cite{capture} (Fig. \ref{results}, top panel). 
(ii) We propose a theoretical explanation for our observations, based on the competition between trapping due to persistent motion and collective expulsion due to the high ``energy'' of tilt walls \cite{Williams} in the trapped aggregate \cite{peruani}. Rotational diffusion, which opposes condensation in motility-induced phase separation \cite{MIPS}, is rendered ineffective in aggregates of elongated particles as explained below. (iii) Our theory predicts, and experiments confirm, that trapping is all-or-nothing: a trap fills to the brim if $\theta < \theta_c$, and particles always escape if $\theta > \theta_c$. A further prediction, that $\theta_c$ goes down with increasing rotational noise, is confirmed in our simulations. (iv) Placed amidst a homogeneous bidisperse mixture, the trap rejects particles with noisy motion and collects those with persistent motility (Fig. \ref{results}, bottom panel). 

The rest of this paper is organized as follows. In Sec. \ref{sec:expsim} we describe our experiments and simulations of trapping. In Sec. \ref{sec:theory} we present a theory of the trapping of active polar rodlike particles. Sec. \ref{sec:sorting} presents our results on the sorting, and we end with a brief conclusion in \ref{sec:conclusion}. 
 
\section{Experiments and Simulations} \label{sec:expsim}
\subsection{Technical details} \label{subsec:techdetail}  
We now describe our findings in detail. Our experiments are carried out in 
a shallow circular geometry with 13 cm diameter {\cite{VJScience,NKPRL}} and a scalloped boundary in order to avoid clustering {\cite{NatCom,dauchotPRL}}. It is covered by a 
glass lid at 1.2 mm above the surface, thus forming a confined two-dimensional 
system. We work with geometrically polar brass rods of length $\ell = 4.5$ mm 
and diameter 1.1 mm at the thick end [Fig. \ref{results}(e)]. The cell is fixed on a permanent-magnet 
shaker (LDS V406-PA 100E) which drives the plate sinusoidally in the vertical 
direction with amplitude $a_{0}$ and frequency $f$= 200 Hz, corresponding to 
dimensionless shaking strength $\Gamma \equiv {a_{0}(2 \pi f)^{2}}/{g} = 7.0$ , 
where $g$ is the acceleration due to gravity. The rod transduces the vibration 
into predominantly forward motion in the direction of its pointed end 
(Supplemental Material video 1 \cite{movies}).  A high-speed camera (Redlake
MotionPro X3) records the dynamics of the particles and ImageJ \cite{ImageJ} is used to extract instantaneous position, orientation and  velocity of the rods.

We use V-shaped traps of aluminium with arm length $L = 10 \ell \simeq 4.5$ cm 
with $20^{\circ} \leq \theta \leq 160^{\circ}$ in steps of $10^{\circ}$. 
The trap is placed in the middle of the cell and stuck to the surface with 
double-sided tape. The cell is filled by a layer of rods spread homogeneously 
and isotropically on the surface. All experiments on the trapping 
in monodisperse systems are done with the number of rods fixed at 150.

Mechanically faithful simulations, with details as in \cite{NatCom}, are 
conducted to investigate which properties of individual particles govern 
their propensity to get trapped and, hence, which features 
control activity-based sorting. 
We construct the tapered rods as arrays of overlapping spheres of 
different sizes \cite{NatCom}. Vibrating base and lid are represented by two 
hard horizontal walls whose vertical ($z$) positions at time $t$ are 
$\mathcal{A} \cos 2 \pi f t$ and $\mathcal{A} \cos 2\pi f t + w$, respectively. In our simulations we work at $\phi_r/F = 1.2$, where 
$\phi_r$ is rod area fraction and $F$ is the ratio of trap area $A_t = L^2 \sin 
\theta/2$ to that of the base, to ensure that the trap does not trivially 
exhaust the total number of rods available on the base plate. We choose periodic boundary conditions 
in the $xy$ plane with linear dimensions of $19.3 \ell$, consistent with the 
experimental geometry. We also run simulations at larger system sizes, always at 
$\phi_r/F = 1.2$. Surface imperfections cause the rods in the experiments to 
perform rotational diffusion, which we capture in the 
simulation \cite{NatCom} by supplying a random angular velocity, 
{$\omega_{z}=\varepsilon v_{rel}$ where $\varepsilon = 
\pm 0.$1 cm$^{-1}$ with equal probability, whenever a rod collides with the base or the 
lid with relative velocity $v_{rel}$ of contact points normal to {the} contact plane, 
and the value 0.1 is chosen to reproduce the {experimentally} observed orientational diffusion. 
We set the values of friction and 
restitution coefficients $\mu$ and $e$ to  0.05 and 0.3 for particle-particle 
collisions, 0.03 and 0.1 for rod-base collisions, 0.01 and 0.1 for rod-lid 
collisions, and 0.03 and 0.65 for particle-V collisions, 
respectively, to match the experiments as best we can. The ballistic dynamics 
of the particles is governed by Newtonian rigid body dynamics. VMD software 
\cite{VMD} is used to make all movies and snapshots from simulations.

\subsection{Results} \label{subsec:res} 
Figures \ref{results}(a) and \ref{results}(c) show the configuration of rods below and above the 
critical angle in experiments (see Supplemental Material video 2 \cite{movies}).  Figures \ref{results}(b) and \ref{results}(d) 
show the corresponding picture in simulation, with system size of $39\ell$ 
(Supplemental Material video 3 \cite{movies}).

\begin{figure}
	\begin{center}
		\includegraphics[width=0.48\textwidth]{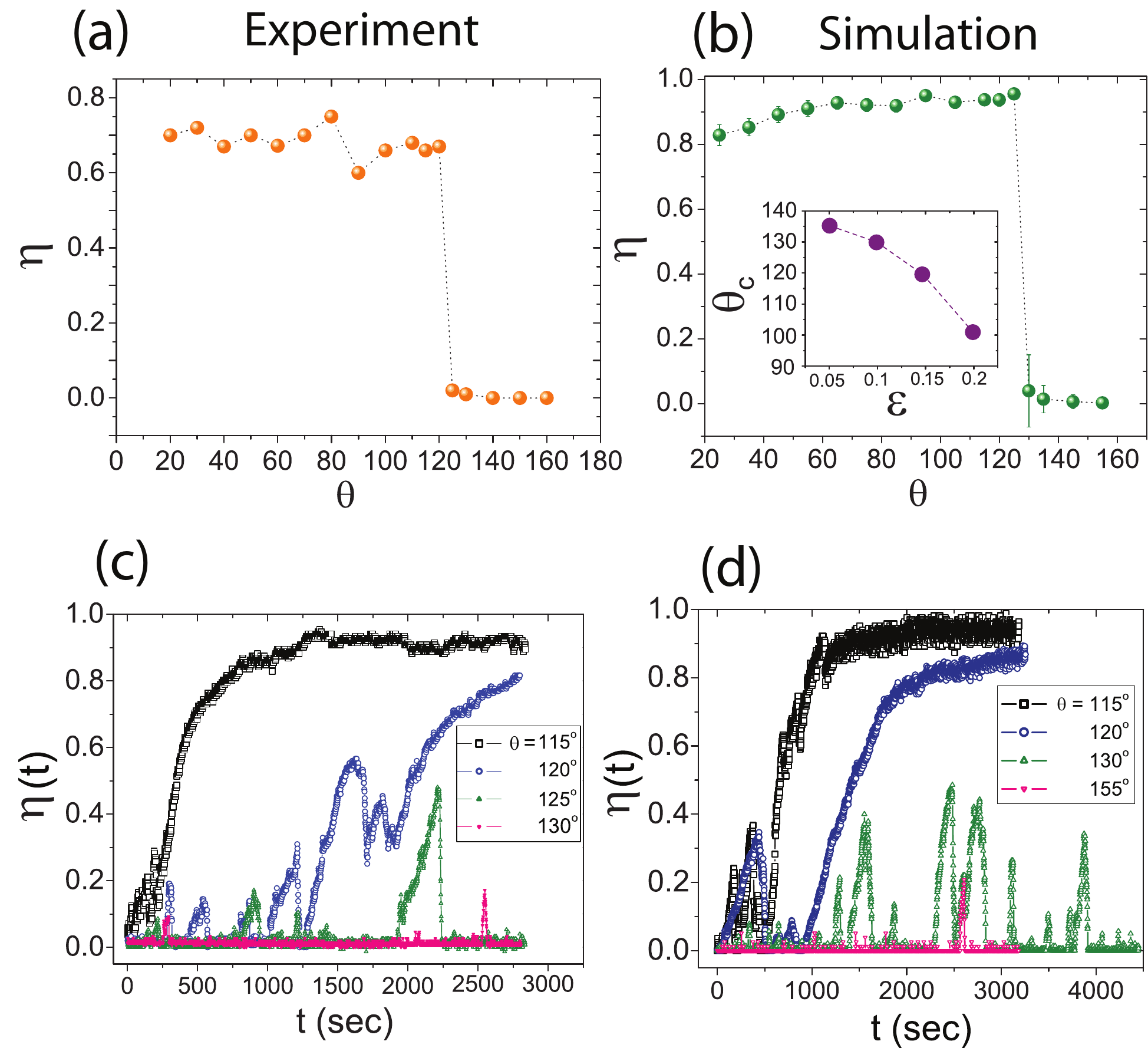}\\
		\caption{Trapping efficiency shows a sudden jump at $\theta = 
120\deg$ for (a) experiment and (b) simulation, indicating a trapping to 
detrapping
			transition. The critical transition angle decreases monotonically with angular noise as shown in the inset to (b). Close to the transition angle, we see repeated attempts where rods tend to form metastable structures
			inside the trap which become increasingly rare as we 
move away from the transition angle, for both experiment (c) and 
simulation (d).}
		\label{transition}
	\end{center}
\end{figure}

In order to quantify trapping, we calculate the instantaneous speeds of all the 
particles in every frame and track the number $N_0$ at zero velocity. We 
plot the trapping efficiency $\eta \equiv N_0 a / A_t$, where 
$a$ is the area of the two-dimensional projection of the rod on the 
surface. We find that 
this quantity drops abruptly at $\theta = 120\deg$ in experiment and 
$\theta = 125\deg$ in the simulation [Fig \ref{transition}(a) and \ref{transition}(b), 
respectively]. We study the effect of angular noise on the threshold 
angle $\theta_c$ below which trapping occurs. We find a monotonic decrease of 
$\theta_c$ with increasing angular noise $\varepsilon$ [see inset to Fig. 
\ref{transition}(b)]. Beyond $\varepsilon = 0.2$ cm$^{-1}$ it is hard to observe 
trapping at any $\theta$. 

We have run longer experiments (45 min) and simulations at trap angles 
$115\deg, 120\deg, 125\deg$ and $130\deg$. We plot the time-evolution of $\eta$ 
in experiment Fig. \ref{transition}(c) and simulation Fig. \ref{transition}(d). 
For both experiment and simulation, with $\theta = 115\deg$, $\eta$ 
increases monotonically with time and saturates to a value 
close to 0.9. At $120\deg$ in the experiment the system displays multiple 
attempts, one of which leads to trapping (see Supplemental Material video 4 from the experiments for an example \cite{movies}). At somewhat larger 
$\theta$ in simulation and experiment such peaks in $\eta(t)$ are seen but are 
ultimately unsuccessful.\\\\

\section{Theory of the trapping transition of active polar rods} \label{sec:theory}
We now show that trapping can be understood theoretically by a nontrivial modification of the idea of motility-induced phase separation (MIPS) \cite{MIPS}. Consider isotropic but self-propelled particles of size $w$, at bulk concentration $c$, each moving with speed $v$ in an independent direction which diffuses rotationally with persistence time $\tau$. The surface of an incipient aggregate will see an incoming current density $\sim cv$, and escape by rotational diffusion at rate $1/\tau$ per particle. The net flux into an incipient aggregate surface is thus $cv - (w^{d-1} \tau)^{-1}$, so the aggregate grows, i.e., MIPS takes place, if the bulk packing fraction $c w^d \equiv \phi>\phi_c \equiv w /v \tau$ \cite{MIPS}.

The trap, the shape of the particles, and the consequent structural order at high packing fraction radically alter the above isotropic bulk MIPS scenario. Although influx is still governed by persistent self-propelled motion,
 
mutual steric hindrance of the anisotropic particles effectively rules out rotational diffusion as a means of escape from an existing aggregate. Why then is there a threshold $\theta$ above which trapping is ineffective? The answer lies in examining the structure of the trapped aggregate.
\begin{figure}
	\begin{center}
		\includegraphics[width=0.48\textwidth]{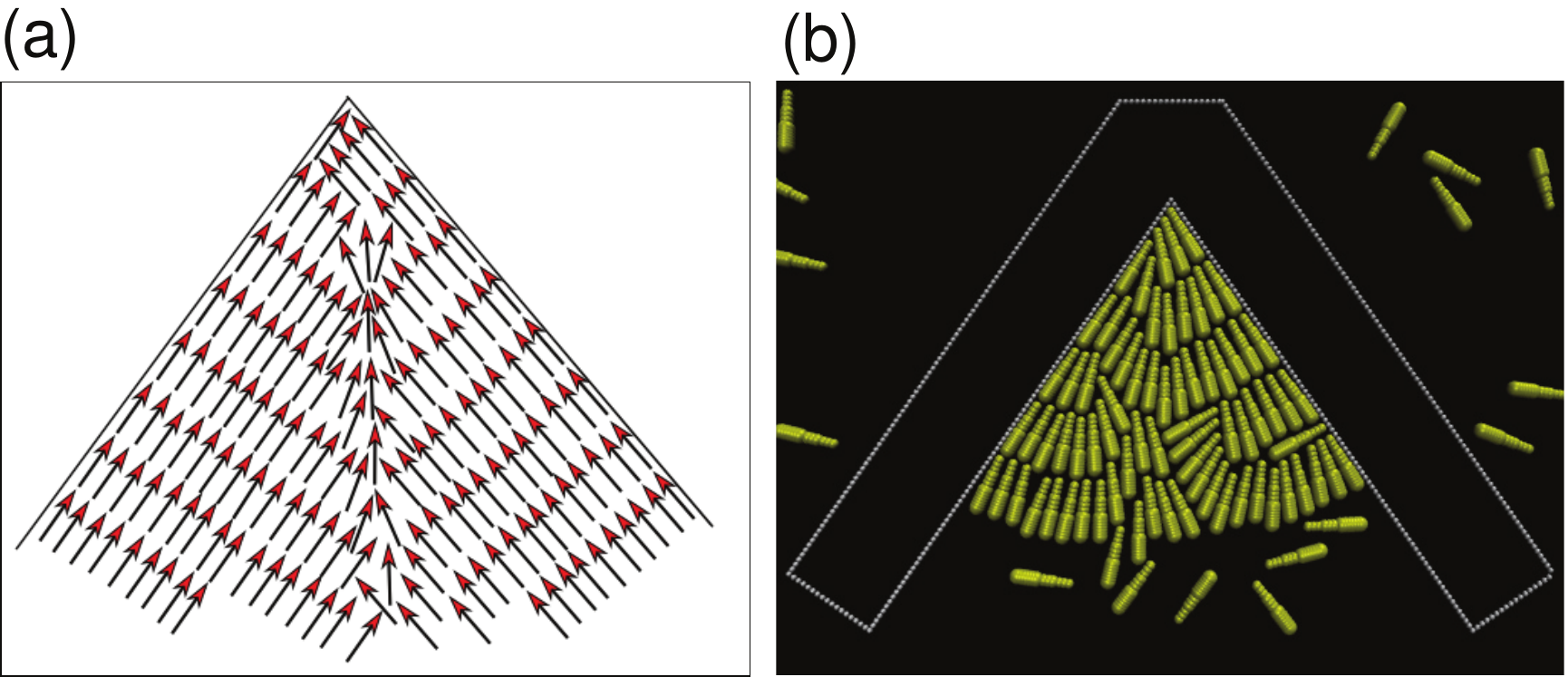}\\
		\caption{(a) Idealized schematic representation of smectic ordering inside a wedge, with a tilt boundary line down the center. (b) Ordering of particles inside a wedge as seen in simulation ($\varepsilon$ = 0.2 cm$^{-1}$).}
		\label{smectic}
	\end{center}
\end{figure}
The elongated shape of the particles leads to strongly anisotropic ordering of the dense trapped aggregate. Our observations strongly suggest that this ordering is not only orientational but translational as well, and in fact closely resembles that of a smectic liquid crystal [see, e.g., Figs. \ref{smectic}(b) and \ref{results}(a)]. Smectic structures are especially evident in \cite{capture}, and in the simulations of \cite{peruani}, who also argue for a defect-based instability of active-rod aggregates but do not present a theoretical calculation and do not appear to consider \textit{translational} order.
 The physics of a smectic in a wedge \cite{Williams} involves the interplay of alignment with the arms of the wedge and the need to preserve the layer spacing. At each arm of the wedge, the normal to the layers is parallel to the arm, with the layer preserving its orientation up to the bisector of the wedge. There is thus [see schematic Fig. \ref{smectic}(a)] an inevitable zone of layer dilation down the middle of the wedge, partially relieved by dislocations if the wedge angle is large \cite{Williams}. For a two-dimensional sample of smectic liquid crystal of radial extent $r$, this can be seen \cite{Williams} to lead to an energy per unit length proportional to the linear extent $r$ of the trapped aggregate, times an increasing function of $\theta$: $E_{wedge} = G \theta^{\alpha} r/w$ where $G$ is a prefactor with units of energy, proportional to the smectic layer-compression modulus, and $\alpha = 2$ ($3$) with (without) the inclusion of dislocations. Adapting this idea to the present case (although we do not have an estimate of the effective elastic constant and hence of $G$), we argue that there is an restoring force $-\partial_r E_{wedge}$ favoring the reduction of the size $r$ of the aggregate, and hence an $r$ velocity given by a mobility times this force:

\begin{equation} 
	\label{wedge_vel}
\left({dr \over dt}\right)_{wedge} = -M {w \over r} {\partial E_{wedge} \over \partial r} = -{M  G\theta^\alpha \over r} = - MG {\theta^{\alpha} \over r}
\end{equation} 
where we have written the mobility in the form $M w/r$ to emphasize that it is the inverse of the drag coefficient of the boundary of the aggregate. The drag should be proportional to the linear size $r$ of the boundary, hence mobility $\sim 1/r$. Thus the wedge energy tends to make the aggregate shrink at a rate $-MG \theta^{\alpha}/r = -D_{eff} \theta^{\alpha}/r$ where $D_{eff}=MG$ is an effective translational diffusivity for $r$. Although we do not have an estimate of $M$ and $G$ individually, we will use a rough measure of the translational diffusivity of the rods, about $10^{-3}$ cm$^2$/s  calculated as given in \cite{DJDurian}, based on displacement along the length of a rod measured on the shortest accessible timescale, i.e., between successive frames \cite{NKPRE}.

\begin{figure}
	\begin{center}
		\includegraphics[width=0.3\textwidth]{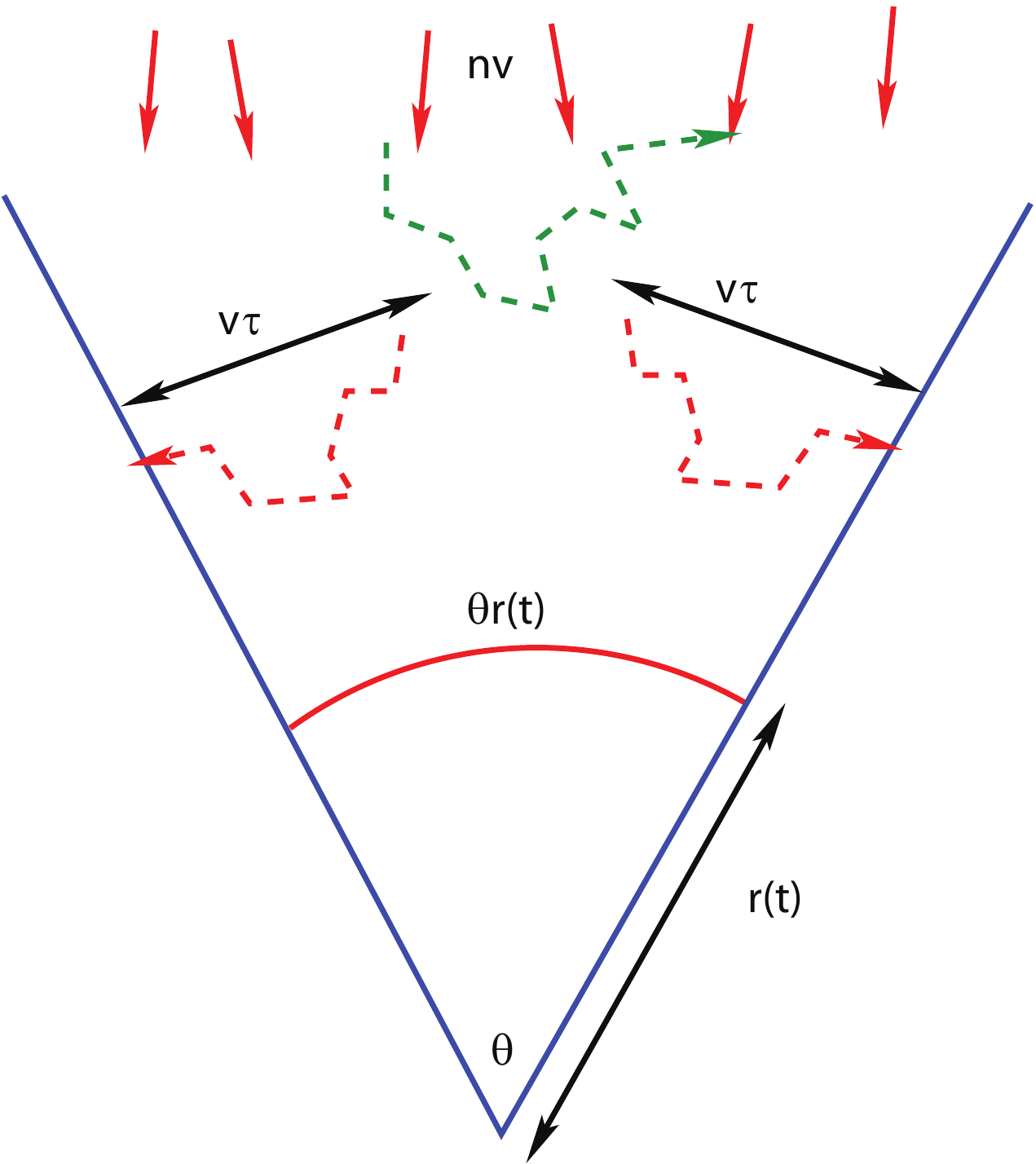}\\
		\caption{Active polar particles at number density $n$ and self-propelled speed $v$ produce a flux of order $nv$ at the mouth of the trap. Admission to the trap (red dashed lines) is primarily for particles captured when they venture within a persistence length $v \tau$ of the arms. Particles entering far from the arms typically escape (green dashed line) by rotational diffusion of their axis, which is also the direction of their self-propelled velocity. An aggregate of length $r$ and width $\sim r \theta$ forms in the trap.}
		\label{trapzone}
	\end{center}
\end{figure}

We must now compare the escape rate \eqref{wedge_vel} with the capture rate. The latter is dominated by 
particles entering within a distance $\sim v {\tau}$ from either arm of the trap, as their escape by rotational diffusion is blocked by collision with the arm (see Fig. \ref{trapzone}). In two dimensions the rate of growth of the aggregate due to influx of particles is then flux $\times$ capture-zone width $\times$ area of particle $\div$ width of aggregate $= c v \times v {\tau} \times w^2/ r \theta = \phi v^2 \tau / r\theta$ where $\phi = c w^2$ is the packing fraction in the bulk, and we have ignored the particle shape in estimating its area as $w^2$. Combining capture and collective escape gives 
\begin{equation} 
\label{totalvel}
{dr \over dt} = \phi v {v {\tau} \over r \theta} - D_{eff} {\theta^{\alpha} \over r}.
\end{equation}
Interestingly both growth and shrinkage are due to terms of order $1/r$ so the parameter values control which wins. As claimed at the start of this paper, and as confirmed within reason in our experiments and simulations, Eq. \eqref{totalvel} says that trapping is all-or-nothing: for a given $\phi$ the aggregate grows without bound if $\theta < \theta_c \simeq (\phi v^2 \tau / D_{eff})^{1/(1 + \alpha)}$, and shrinks to nothing for $\theta > \theta_c$. For the system at hand $\phi \simeq 0.1$, $v \simeq 0.34$ cm/s \cite{NKPRE}, and $\tau \simeq 0.16$ s, so that $\theta_c \simeq 2^{1/(1 + \alpha)}$, which is about a radian. These are rough estimates, up to unknown prefactors of order unity, so the factor of 2 with respect to the observed value of about 120$^{\circ}$ is tolerable. Moreover, the prediction that $\theta_c$ decreases with increasing noise strength is in agreement with computer experiments in which the angular noise was varied [see inset to Fig. \ref{transition}(b)]. 

Limitations of finite size complicate a detailed comparison of theory to experiment. The assumed picture of a smectic in a wedge requires an aggregate large in length $r$ and width $r \theta$, i.e., progressively larger $r$ as $\theta$ is decreased. Estimating the threshold $\theta_c$ as the angle at which $dr/dt$ vanishes assumes that the trap has room to absorb more particles and that an ample supply of particles is present in the ambient medium. Future experimental tests of \eqref{totalvel}, whether laboratory or numerical, will need to keep these constraints in mind.

\section{Sorting active particles: an application of the trapping mechanism} \label{sec:sorting} 
It was remarked in Ref. \cite{capture} that increasing angular noise favors escape from the trap. Prompted by this observation, we examine here the possibility of sorting based purely on the statistical character of active motion. This exploration is complementary to the many studies of sorting by motility type \cite{sort}. In our experiments, the noise has translational and rotational contributions, and depends in detail on particle shape. 
\begin{figure}[!t]
\begin{center}
\includegraphics[width=0.47\textwidth]{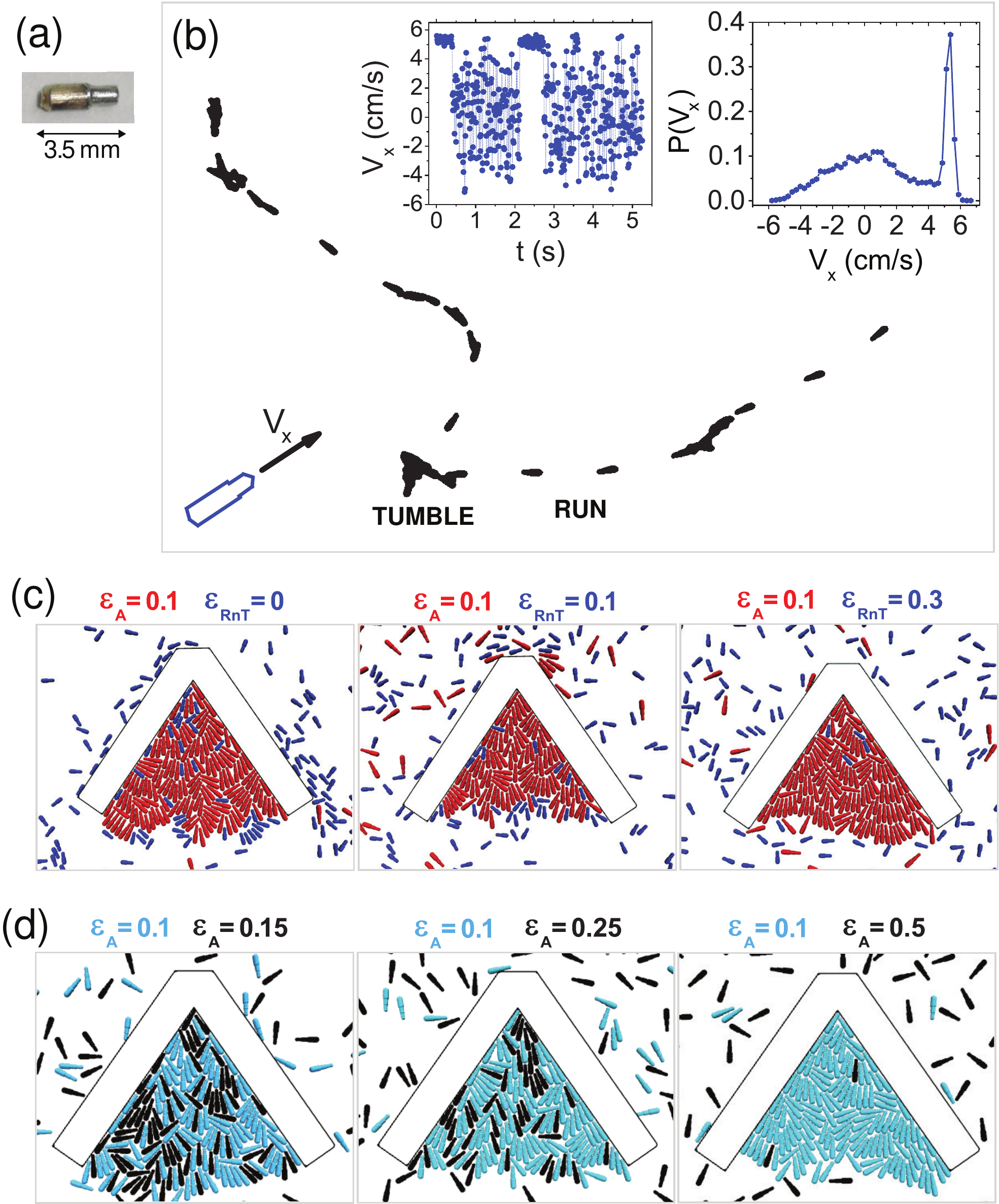}\\
\caption{(a) A photograph of the  R{\&}T particle along with its x-component of the in-plane velocity $V_x$ and orientation $\theta$. (b) A typical trajectory of the particle showing run and tumble events with time. Plot of $V_x$ as function of time and its probability distribution in the inset. (c) Steady states of mixture of A and  R{\&}T for constant $\epsilon_A$ and different $\epsilon_{ R{\&}T}$, showing insensitivity of R{\&}T particles to added angular noise. (d) Steady states of binary mixtures of type A particles distinguished only by angular noise strength; sorting sets in with increasing noise contrast.}
\label{segg}
\end{center}
\end{figure}
To study sorting, we introduce a second type of active polar rod [see Fig. \ref{segg}(a)], 3.5 mm long and 1.1 mm thick at its thicker end, and tapered in a 
single step, with dynamics qualitatively different from that 
of the particles (hereafter type A) discussed in the first part of this paper. 
Fig. \ref{segg}(b) shows a typical trajectory of this particle (see 
Supplemental Material video 5 \cite{movies}), displaying rapid directed \textit{runs} interrupted by 
abrupt \textit{tumbles} following which a new run direction is selected at 
random. In what follows we will refer to these as  R{\&}T particles although, 
unlike in bacteria \cite{ecoli}, the tumbles here are 
generally longer than the runs. The insets to Fig. \ref{segg}(b) show a 
typical time-trace of $V_{x}$, the instantaneous velocity component along the 
long axis, and the probability distribution $P(V_x)$. Note the strongly bimodal 
character of $P(V_x)$, with a sharp peak corresponding to the run motion. 
Experiments similar to those discussed above show that our  R{\&}T particles are 
not trapped for any $\theta$. 

We now introduce an initially homogeneous mixture of 150 type A and 225  R{\&}T 
particles into the sample cell containing a trap with $\theta = 70\deg$. 
Upon vertical shaking we find a strongly selective trapping of only the A 
particles, with  R{\&}T entering and leaving freely. The four images in time 
sequence in Fig. \ref{results}(e) illustrate this separation, and the 
Supplemental Material video 6 \cite{movies} shows the kinetics in detail. 

In order to study what aspect of shape or kinetics governs the relative 
susceptibility to trapping, we carry out simulations in which we have 
independent control over particle properties. We do not attempt to recreate the 
complex dynamics of the  R{\&}T particles. We retain the shapes of the A and  R{\&}T 
particles, but force them with angular white noise. We keep the A particle 
noise at levels consistent with the experiment, and vary the noise strength on 
the  R{\&}T particles. As seen in Fig. \ref{segg}(c) (and Supplemental Material video 7 \cite{movies}) for zero noise, intermediate noise and high noise respectively, the sorting is highly effective in all 
cases, with no perceptible effect due to the angular noise on the  R{\&}T.
This is presumably a consequence of the sensitivity to initial 
conditions of their deterministic dynamics, through the interplay of 
agitation, shape and interaction with the bounding surfaces. This 
lack of persistence is what saves the  R{\&}T particles from being trapped, reminiscent of reversal behaviour in bacteria \cite{myxo}, albeit with a different mechanism. Lastly, we study mixtures of geometrically identical A particles, distinguished only by an imposed difference in their angular noise. Again, Fig. \ref{segg}(d) (and Supplemental Material video 8 \cite{movies}), in increasing order of difference in noise strengths, show highly effective sorting at large noise difference, with the 
trap predominantly populated by the less noisy, more persistent component. 

Thus, the trappability of a particle is linked to the persistence of its 
directed motion. Reducing this persistence, whether through angular noise or 
enhanced shuffling along the axis of the particle, facilitates escape from the 
trap, and results in a preferential accumulation of persistent movers inside 
the trap.\\\\

\section{Conclusion} \label{sec:conclusion} 
In summary, our experiments and simulations find a phase transition to a collectively trapped state when a V-shaped obstacle is introduced amidst a monolayer of artificially motile macroscopic rods, as the angle of the V is decreased past a threshold $\theta_c$. We offer a theory of this transition based on the competition between accumulation inside the V due to persistent motion, and expulsion due to costly tilt walls in the layered structure formed within the trap. The theory predicts that trapping is all-or-nothing, in agreement with experiments, and that $\theta_c$ decreases with increasing noise strength, confirmed in our simulations. Crude parameter estimates yield $\theta_c$ within a factor of 2 of the value observed. Finally, we demonstrate sorting: from a mixture, the trap spontaneously gathers persistent movers and excludes noisy particles. Our results are a key step toward a complete theory of MIPS \cite{MIPS} in polar active liquid crystals \cite{SriramAnnRev,SriramRMP}. 

\section{Acknowledgements} We thank the University Grants Commission (N.K.); the Council of Scientific and Industrial Research (HS); the DST, India (A.K.S., support from a Year of Science Professorship); the Science and Engineering Research Board, India (S.R., support from a J C Bose Fellowship); the Tata Education and Development Trust (S.R.); and the Department of Physics, Indian Institute of Science (R.K.G., for hospitality and support).

\end{document}